\documentclass[aps,reprint,amsmath,amssymb,superscriptaddress,preprintnumbers,nofootinbib]{revtex4-2}

\usepackage{graphicx}
\usepackage{dcolumn}
\usepackage{bm}
\usepackage{braket}
\usepackage{hyperref}
\usepackage[export]{adjustbox}
\usepackage{mathtools}
\usepackage{dcolumn}
\usepackage{siunitx}

\usepackage{subfigure}

\usepackage{dsfont}

\usepackage{color}
\usepackage[dvipsnames]{xcolor}

\newcommand{\eg}{\emph{e.g.}~}

\newcommand{\beq}{\begin{equation}}
\newcommand{\eeq}{\end{equation}}
\newcommand{\bea}{\begin{eqnarray}}
\newcommand{\eea}{\end{eqnarray}}

\newcommand{\Lag}{\mathcal{L}}

\newcommand{\Mpl}{M_{\text{Pl}}}
\newcommand{\dmi}{d_{m_i}^{(2)}}

\newcommand{\dme}{d_{m_e}^{(2)}}

\newcommand{\de}{d_{e}^{(2)}}

\newcommand{\dg}{d_{g}^{(2)}}

\DeclareRobustCommand{\Sec}[1]{Sec.~\ref{#1}}

\DeclareRobustCommand{\App}[1]{App.~\ref{#1}}

\DeclareRobustCommand{\Fig}[1]{Fig.~\ref{#1}}

\DeclareRobustCommand{\Eq}[1]{Eq.~(\ref{#1})}

\newcommand{\iso}[2]{{\ensuremath{{}^{#2}}\ensuremath{\rm #1}}}

\usepackage{verbatim}

\definecolor{Std1}{rgb}{0.24, 0.6, 0.8}
\definecolor{Std2}{rgb}{0.95, 0.627, 0.1425}
\definecolor{Std3}{rgb}{0.455, 0.7, 0.21}
\definecolor{Std4}{rgb}{0.922526, 0.385626, 0.209179}
\definecolor{Std5}{rgb}{0.578, 0.51, 0.85}
\definecolor{Std6}{rgb}{0.772079, 0.431554, 0.102387}
\definecolor{Std7}{rgb}{0.4, 0.64, 1.}
\definecolor{Std8}{rgb}{1., 0.75, 0.}
\definecolor{Std9}{rgb}{0.8, 0.4, 0.76}
\definecolor{Std10}{rgb}{0.637, 0.65, 0.}
\definecolor{BBN}{rgb}{0.6, 0., 0.}
\definecolor{Th229}{rgb}{0.333, 0.333, 0.333}

\begin{document}

\title{
Natural Ultralight Dark Matter: The Quadratic Twin
}

\author{C\'edric Delaunay}
\email{cedric.delaunay@lapth.cnrs.fr}
\affiliation{Laboratoire d’Annecy de Physique Th\'eorique, CNRS – USMB, 74940 Annecy, France}

\author{Michael Geller}
\email{micgeller@tauex.tau.ac.il}
\affiliation{School of Physics and Astronomy, Tel-Aviv University, Tel-Aviv 69978, Israel}

\author{Zamir Heller-Algazi}
\email{zamir.heller@gmail.com}
\affiliation{School of Physics and Astronomy, Tel-Aviv University, Tel-Aviv 69978, Israel}

\author{Gilad Perez}
\email{gilad.perez@weizmann.ac.il}
\affiliation{Department of Particle Physics and Astrophysics, Weizmann Institute of Science, Rehovot 7610001, Israel}

\author{Konstantin Springmann}
\email{konstantin.springmann@weizmann.ac.il}
\affiliation{Department of Particle Physics and Astrophysics, Weizmann Institute of Science, Rehovot 7610001, Israel}

\date{\today}

\begin{abstract}
Scalar ultralight dark matter (ULDM) is uniquely accessible to tabletop experiments such as clocks and interferometers, and its search has been the focus of a vast experimental effort. 
However, the scalar ULDM mass is not protected from radiative corrections, and the entirety of the parameter space within reach of experiments suffers from a severe naturalness problem. 
In this paper, we propose a new twin mechanism that protects the mass of the scalar ULDM. 
Our scalar ULDM is a pseudo-Nambu-Goldstone boson with quadratic couplings to the Standard Model (SM) and to a twin copy of the SM, with a mirror $\mathbb{Z}_2$ symmetry exchanging each SM particle with its twin. 
Due to the mirror symmetry, the leading-order mass correction is quadratic in the (tiny) coupling while the linear order is canceled.
This opens up vast regions of parameter space for natural quadratically coupled ultralight dark matter, within the sensitivity of existing and future experiments.
\end{abstract}

\maketitle

\section{Introduction}\label{sec:intro}

The dark matter (DM) problem has been a major driver of research in particle physics, cosmology, and more recently, in atomic, molecular, and optical physics. 
A large part of this endeavor focuses on the exciting possibility that DM consists of an ultralight (or sub-eV) degree of freedom, whose coherent oscillations account for the observed DM density \cite{Preskill:1982cy,Dine:1982ah}. 
This scenario has far-reaching experimental implications, with search strategies ranging from galaxy dynamics to tabletop experiments. For a recent review see \cite{Antypas:2022asj}.

One of the most promising avenues to search for ultralight dark matter (ULDM) is when it has linear or quadratic couplings to CP-even Standard Model (SM) operators. 
In this case, astrophysical objects like Earth can distort the DM background and change its density profile~\cite{Hees:2018fpg,Banerjee:2022sqg,Budker:2023sex,Banerjee:2025dlo,delCastillo:2025rbr}, and so these interactions can be searched for in tests of the equivalence principle (EP)~\cite{Antypas:2022asj}.
Astrophysical objects may trigger a tachyonic instability in the DM field, whose backreaction can lead to dramatic and observable effects on these objects~\cite {Hook:2017psm,Balkin:2020dsr,Balkin:2022qer,Balkin:2023xtr,Gomez-Banon:2024oux,Kumamoto:2024wjd}. 
Furthermore, the coherent DM oscillations induce corresponding oscillations in gauge couplings and particle masses --- an effect that can be probed by clocks~\cite{Antypas:2022asj}, laser excitations of \iso{Th}{229}~\cite{Fuchs:2024edo,Delaunay:2025lgk}, and in the future by a fully operating nuclear clock~\cite{2021NatRP...3..238B} with extraordinary sensitivity~\cite{Caputo:2024doz}.

This possibility, however, is generally plagued by a naturalness problem: ULDM couplings to SM operators typically introduce overly large radiative corrections to the ULDM mass.
As a result, one must either accept small couplings, which render the ULDM invisible, or accept unrealistically low cutoff scales for SM loops.

To illustrate this point, consider an ULDM field $\phi$ with a linear or a quadratic coupling to the electron mass operator, denoted $d_{m_e}^{(1)}/\Mpl$ and $\dme/(2\Mpl^2)$, respectively, where $\Mpl$ is the reduced Planck mass. 
Setting for instance $m_\phi=10^{-15}\,$eV, existing experimental bounds on these couplings are $d_{m_e}^{(1)}\lesssim 10^{-3}$ and $d_{m_e}^{(2)}\lesssim 10^{12}$, respectively, see \eg~\cite{Banerjee:2022sqg}. 
For the linear case, current experiments probe cutoffs of $\mathcal{O}(100)\,$GeV, which would imply new physics at the weak scale that has not been discovered. 
For quadratically coupled ULDM, the situation is even worse since currently the best experiments are sensitive to SM cutoffs of $\mathcal{O}(10)\,$eV.

Such fine-tuning problems are common to many scalar and pseudoscalar ULDM theories \cite{Froggatt:1978nt,Leurer:1992wg,Leurer:1993gy,Nelson:1983zb,Nelson:1984hg,Barr:1984qx,Piazza:2010ye,Graham:2015cka,Calibbi:2016hwq,Dine:2024bxv}, and have inspired model-building solutions for various classes of models.
ULDM theories where the DM field is a pseudo-Nambu-Goldstone boson (pNGB) are often considered in the literature because the pNGB mass is protected by an approximate shift-symmetry.
A prominent model is the QCD axion \cite{Peccei:1977hh,Wilczek:1977pj,Weinberg:1977ma}, which can play the role of the DM and also solve the strong CP problem.
But even with the protection of the approximate shift-symmetry, most experiments are typically not sensitive enough to probe its natural parameter space, and hence require further tuning.
The only natural model \cite{Hook_2018b,Banerjee:2025kov} (see also \cite{DiLuzio:2021pxd}) populating the region currently probed by experiments as they approach the QCD axion line, envokes $N$ copies of the SM that are related by a discrete $\mathbb{Z}_N$ symmetry, which conspire to protect the QCD axion mass.
The same mechanism has also been used for natural ULDM with linear couplings to the SM in \cite{Brzeminski:2020uhm}.
The dilaton has also been proposed as an ULDM candidate~\cite{Cho:1998js,Damour:2010rm,Damour:2010rp,Arvanitaki:2014faa}, but due to its generically large quartic, its early universe behavior is in tension with observations in standard theoretical setups \cite{Hubisz:2024hyz}. 
While a consistent theory of the dilaton ULDM model was proposed in \cite{Banerjee:2025uwn}, its natural parameter space is also well beyond the reach of any experiment.

\begin{figure}[t!] 
  \centering
\includegraphics[width=0.5\textwidth]{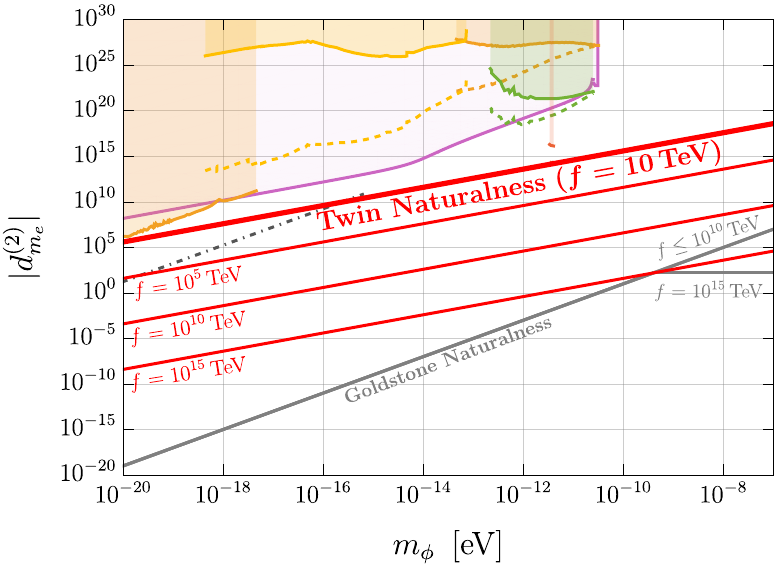}
 \caption[]{
 The parameter space of quadratically coupled ULDM: coupling to electrons $|\dme|$ vs. mass $m_\phi$. 
 The red lines are the naturalness bounds from our quadratic twin ULDM for a cutoff $\Lambda=10\,\text{TeV}$ and various values of $f$.
 For comparison, we plot the Goldstone naturalness bound (gray) for the same cutoff and various values of $f$ without the twin mechanism, see \App{app:naturalness}. 
 DM amplitude-sensitive bounds, subject to screening, are shown in solid lines, while dashed lines of the same color neglect screening \cite{Hees:2018fpg,Banerjee:2025dlo}, see \Sec{sec:Pheno}.
 Among them are atomic clock bounds {\color{Std2}{\textbf{H/Si}}}~\cite{Kennedy:2020bac}, {\color{Std8}{\textbf{quartz}}}~\cite{Campbell:2020fvq}, {\color{Std2}{\textbf{molecular}}}~\cite{Oswald:2021vtc,Oswald:2025bih}, and interferometers {\color{Std3}{\textbf{Geo600}}}~\cite{Vermeulen:2021epa} and {\color{Std4}{\textbf{Auriga}}}~\cite{Branca:2016rez}.
 Additionally, we show gradient-sensitive constraints from {\color{Std9}\textbf{MICROSCOPE}}~\cite{Banerjee:2025dlo,Gue:2025nxq} for $\dme>0$, and the projected sensitivity of the {\color{Th229}{\boldsymbol{\iso{Th}{229}}}} nuclear clock (dot-dashed).}
  \label{fig:parspace}
\end{figure}

This work aims to alleviate the naturalness problem of ULDM theories with quadratic couplings large enough to be detected by ongoing and future experiments. 
Inspired by the twin-Higgs mechanism \cite{Chacko:2005pe}, we consider theories where the ULDM fields are pNGBs whose masses are protected by a mirror-$\mathbb{Z}_2$ symmetry relating the SM sector to a twin sector, which is a copy of the SM. 
As shown in \Fig{fig:parspace}, this theoretical construct opens up roughly 25 orders of magnitude in the parameter space of couplings for DM masses $m_\phi\sim(10^{-20}-10^{-15})\,\text{eV}$, which puts natural models at the doorstep of ongoing and near-future experiments. 

\section{The quadratic twin mechanism}

We illustrate the mechanism at the heart of our proposal by considering, for illustration, the simplest perturbative UV completion, in which the pNGBs reside inside a complex scalar field $\Sigma$ transforming as the fundamental of an approximate global symmetry group $G$ with potential,\footnote{Equivalently, one can describe the same dynamics using a nonlinear sigma model below the scale $f$.}
\begin{equation}
\label{eq:SymBreakPot}
    V(\Sigma)=\lambda\left(|\Sigma|^2-f^2\right)^2,
\end{equation}
which generates a vacuum expectation value (VEV) ${\left\langle|\Sigma|\right\rangle=f}$ that spontaneously breaks $G$.
Additionally, $G$ is explicitly broken to a subgroup $H_\text{SM}\times H_{\text{twin}}\subset G$ by weakly coupling $\Sigma=(\Sigma_\text{SM},\Sigma_\text{twin})^{\operatorname{T}}$ to our SM and the twin sector,
\begin{equation}\label{eq:ExplicitBreaking}
    \Lag_{\rm explicit}\supset c\left(|\Sigma_\text{SM}|^2\mathcal{O}_\text{SM}+|\Sigma_\text{twin}|^2\mathcal{O}_\text{twin}\right),
\end{equation}
where $c$ is a coupling constant and $\mathcal{O}_\text{SM/twin}$ are scalar operators, singlets under $G$,
of our SM sector and its twin, respectively. 
The mirror-$\mathbb{Z}_2$ exchanges SM $\Leftrightarrow$ twin therefore ensuring identical couplings for both terms.

The Goldstones residing in $\Sigma_\text{SM/twin}$ naturally get a potential from radiative corrections, in addition to the bare mass $m_\text{pNGB}$ which we implicitly assume comes from another source of symmetry breaking.
The mirror $\mathbb{Z}_2$-symmetry restricts the form of these radiative corrections,  ameliorating the hierarchy problem associated with the pNGB mass.
For concreteness, consider a coupling to a SM fermion $\psi$ of the form,
\begin{equation}\label{Eq:ElectronMassOperator}
    c|\Sigma_\text{SM}|^2\mathcal{O}_\text{SM}=\frac{d_{m_\psi}^{(2)} }{\Mpl^2}|\Sigma_\text{SM}|^2m_\psi\bar\psi\psi\,.
\end{equation}
In this particular case, $c=d_{m_\psi}^{(2)}/(2\Mpl^2)$, where $d_{m_\psi}^{(2)}$ is a dimensionless constant and $\Mpl$ is the Planck scale.
In the absence of the twin-sector, the interaction in~\Eq{Eq:ElectronMassOperator} leads to a mass correction 
\begin{equation}\label{eq:NaiveMass}
   \text{SM:}\quad \delta m_\text{pNGB}^2
    \simeq\frac{d_{m_\psi}^{(2)} m_\psi^2\Lambda^2}{16\pi^2\Mpl^2},
\end{equation}
where $\Lambda$ is the cutoff of the theory.
The naturalness criterion ${\delta m_\text{pNGB}^2\lesssim m_\text{pNGB}^2}$, combined with current experimental sensitivity on $d_{m_\psi}^{(2)}$, constrains $\Lambda$ to small scales, \eg of the order $\Lambda\sim10\,\text{eV}$ for coupling to electrons. 
We refer to this as the ``Goldstone naturalness'' criterion; see \App{app:naturalness} for more details.

This contribution is canceled exactly by the twin-sector loop thanks to the mirror-$\mathbb{Z}_2$ symmetry, since the full induced potential,
\begin{equation}\label{eq:twin_linear_correction}
    \Delta V=\frac{d_{m_\psi}^{(2)}m_\psi^2\Lambda^2}{16\pi^2 \Mpl^2}\left(|\Sigma_\text{SM}|^2+|\Sigma_\text{twin}|^2\right)=\frac{d_{m_\psi}^{(2)}m_\psi^2\Lambda^2}{16\pi^2\Mpl^2 } f^2,
\end{equation}
is invariant under the symmetry group $G$, and therefore does not contribute to the pNGB mass.
However, the mirror-$\mathbb Z_2$ symmetry does not cancel the one-loop correction that is quadratic in the coupling. The resulting potential is
\begin{equation}
    \Delta V=-\frac{(d_{m_\psi}^{(2)})^2 m_\psi^2\Lambda^2}{16\pi^2\Mpl^4}\left( |\Sigma_\text{SM}|^4+|\Sigma_\text{twin}|^4\right),
\end{equation}
which implies a mass correction for the pNGB
\begin{equation}
    \text{SM+twin:}\quad\delta m_\text{pNGB}^2
    =\frac{(d_{m_\psi}^{(2)})^2 m_\psi^2 f^2}{16\pi^2\Mpl^4}\Lambda^2.
\end{equation}
While the degree of divergence remains the same\footnote{This is unlike the twin-Higgs mechanism \cite{Chacko:2005pe}, where the explicit breaking arises from renormalizable interactions, and the mass correction is $\log$-divergent instead.} compared to the case without twin contributions in \Eq{eq:NaiveMass}, the mass correction is now quadratic in the coupling. 
Since the pNGB is coupled extremely weakly to the SM, this correction is typically many orders of magnitude smaller than the contribution in
\Eq{eq:NaiveMass}.

The same mechanism works for any symmetry group $G$, including non-abelian groups, and applies to other quadratic couplings, given by the Lagrangian
\begin{equation}\label{eq:FullLag}
\begin{aligned}
      \Lag&=\frac{|\Sigma_\text{SM}|^2}{2\Mpl^2}\bigg[\frac{\de}{4e^2}F_{\mu\nu}F^{\mu\nu}-\frac{\dg\beta_g}{2g}G_{a,\mu\nu}G^{a,\mu\nu}\bigg.\\
      &\bigg.-\left(\dmi+\gamma_{m_i}\dg\right)m_i\bar{\psi}_i\psi_i+\text{h.c.}\bigg]+\text{twin-sector}.
\end{aligned}
\end{equation}
$F_{\mu\nu}$, $G_{a,\mu\nu}$ are the photon and gluon field strengths respectively, $\beta_g$ is the beta-function of the strong coupling constant $g$, $e$ the electromagnetic coupling constant, $\gamma_{m_i}$ the anomalous dimension and $\psi_i$ are the SM fermions. 
We parameterize the dimensionless 
couplings $\de,\dg,\dmi$ following the convention of \cite{Damour:2010rp}.  
The suppression of the pNGB mass correction in our model, relative to the SM contribution is therefore
\begin{equation}\label{eq:UniversalSuppression}
    \sum_id_i\left(\frac{f}{\Mpl}\right)^2\ll1,
\end{equation}
where $i$ runs over the couplings in \Eq{eq:FullLag} and $d_e\equiv\de/4e^2,d_g\equiv\dg\beta_g/(2g)$, etc.

We move on to discuss the scales and couplings in our problem.
Due to the extra insertion of the explicit breaking of $G$ in \Eq{eq:ExplicitBreaking}, the symmetry breaking scale $f$ does not determine the couplings to SM operators.
Consistency of the nonlinear sigma model imposes $f\gtrsim\Lambda$ and we require $\Lambda\gtrsim10\,\text{TeV}$ to avoid collider constraints.
We also require the coupling of $\Sigma_\text{SM}$ to the SM to be technically natural such that $d_i^{1/2}/\Mpl\lesssim 1/f$, see also \App{app:naturalness}.

Finally, we compare the quadratic twin ULDM to theories of QCD axions \cite{Peccei:1977hh,Wilczek:1977pj,Weinberg:1977ma}.
The couplings of the quadratic twin ULDM to the SM and twin sector explicitly break the symmetry group $G$, and are not necessarily linked to any chiral anomaly.
In contrast, the QCD axion necessarily breaks the PQ symmetry by the anomaly, which induces similar quadratic couplings to SM particles and the axion mass (see \eg \cite{Balkin:2020dsr,Kim:2022ype,Kim:2023pvt,Beadle:2023flm,Springmann:2024mjp}). 
Notably, the QCD axion also has derivative couplings from either the anomaly (as in KSVZ models \cite{Kim:1979if,Shifman:1979if}) or the UV (as in DFSZ models \cite{Dine:1981rt,Zhitnitsky:1980tq}). 
For the quadratic twin, no derivative couplings are generated if the explicit breaking is not related to a chiral anomaly.

\section{\boldmath $U(1)$ model}\label{sec:U1Model}

We explicitly demonstrate how our mechanism works for the minimal case of $G=U(1)$. 
Consider a complex scalar $\Sigma$ charged under $U(1)$ with a potential given by \Eq{eq:SymBreakPot}. $\Sigma$ obtains a VEV $\langle |\Sigma|\rangle=f$ that spontaneously breaks the $U(1)$. We parameterize excitations around the vacuum as
\begin{equation}
    \Sigma=f\exp\left({i\frac{\phi}{f}}\right),
\end{equation}
with $\phi$ the Goldstone boson. 
The global $U(1)$ is explicitly broken to the subgroup $H_\text{SM}\times H_{\text{twin}}=\left(\mathbb{Z}_2\right)_\text{SM}\times\left(\mathbb{Z}_2\right)_\text{twin}$, by coupling to SM and twin operators as in \Eq{eq:ExplicitBreaking}.
We parameterize the components of $\Sigma$ transforming under $\left(\mathbb{Z}_2\right)_\text{SM}\times\left(\mathbb{Z}_2\right)_\text{twin}$ as
\begin{equation}\label{eq:U1embedding}
    \Sigma_\text{SM}=f\sin\left(\frac{\phi}{f}\right),\quad\Sigma_\text{twin}=f\cos\left(\frac{\phi}{f}\right).
\end{equation}

We assume that interactions between the Goldstone $\phi$ and SM fields, like the electron, explicitly break the global $U(1)$ symmetry as in \Eq{eq:ExplicitBreaking}. Because the SM fields are not charged under the $U(1)$ there are no linear couplings, so the simplest interaction between $\phi$ and the SM electron $e$ and the twin electron $e'$ is quadratic and of the form
\begin{equation}\label{eq:ULDM_electron_interaction}
\begin{aligned}
    \Lag&\supset-\frac{\dme m_e}{2M_{\text{Pl}}^2}\left(|\Sigma_\text{SM}|^2\bar{e}e+|\Sigma_\text{twin}|^2\bar{e}'e'\right)\\
    &=-\frac{\dme m_e}{2\Mpl^2}\left( \bar{e}e-\bar{e}'e'\right)\phi^2+\dots\,,
    \end{aligned}
\end{equation}
where $\dots$ denotes terms of higher order in $\phi$.
By virtue of the mirror-$\mathbb{Z}_2$ symmetry, both SM and twin electrons have identical coupling $\dme$ to $\phi^2$, while the opposite sign arises due to the embedding in \Eq{eq:U1embedding}.

The quadratic twin ULDM obtains a potential due to loop corrections from \Eq{eq:ULDM_electron_interaction}. 
Because $e$ and $e'$ couple to $\phi^2$ with opposite signs, their loop corrections cancel at linear order in $\dme$. 
At second order in the coupling, we find that the leading order correction to the one-loop potential is
\begin{equation}
\begin{aligned}
    \Delta V(\phi)&=-\frac{(\dme)^2 m_e^2\Lambda^2}{64\pi^2\Mpl^4}\left[ |\Sigma_\text{SM}|^4+|\Sigma_\text{twin}|^4\right]\\
    &=-\frac{(\dme)^2m_e^2f^4\Lambda^2}{256\pi^2\Mpl^4}\left[\cos\left(4\frac{\phi}{f}\right)-1\right],
\end{aligned}
\end{equation}
up to an additive constant. 
The resulting mass correction is
\begin{equation}\label{eq:NatMassQuadTwinULDM}
    \delta m_\phi^2=\frac{(\dme)^2m_e^2 f^2}{16\pi^2\Mpl^4}\Lambda^2,
\end{equation}
which is suppressed compared to the case without the mirror symmetry, as in \Eq{eq:UniversalSuppression}, where $d_{m_e}=\dme$.

We show the parameter space for the absolute value of the coupling in \Fig{fig:parspace}, where we compare the Goldstone naturalness bound (gray) to the twin naturalness bound (red) for ${\Lambda\simeq10\,\text{TeV}}$.
We discuss differences that arise due to the sign of $\dme$ in \Sec{sec:Pheno} and in \App{app:ScreenSource}.

\section{Non-abelian groups}\label{sec:NonAbelianModel}

Our model can be straightforwardly extended to non-abelian groups.\footnote{Interestingly, a similar scenario was recently considered in \cite{Bigaran:2025uzn} and shown to potentially give rise to oscillating exotic meson decays.} 
Here we focus on the complex field $\Sigma$ in the fundamental of $G=U(2)$.
The $U(2)$ is spontaneously broken to a $U(1)$ subgroup by the VEV $\left\langle|\Sigma|^2\right\rangle=f^2$ and explicitly broken to $H_\text{SM}\times H_\text{twin}=U(1)_\text{SM}\times U(1)_\text{twin}$ by couplings to the SM and twin sectors.
This symmetry-breaking pattern leads to 3 pNGBs, which we parameterize as
\begin{equation}\label{eq:U2param}
\begin{aligned}
    \Sigma&=\begin{pmatrix}
    \Sigma_\text{SM}\\
    \Sigma_\text{twin}
\end{pmatrix}=e^{2i\phi^{\hat{a}}T^{\hat{a}}/f}\begin{pmatrix}
        0\\
        f
    \end{pmatrix}\\&=f\frac{\sin(\phi/f)}{\phi}\begin{pmatrix}
    \sqrt{2}i\phi^+\\
    \phi \cot(\phi/f)-i \phi^0
    \end{pmatrix},
\end{aligned}
\end{equation}
where $T^{\hat{a}}\in SU(2)$ with $\hat{a}=1,2,3$ are the broken generators with associated pNGBs $\phi^{\hat{a}}$, which we write as $\phi^0=\phi^3$, $\phi^\pm=(\phi^1\mp i\phi^2)/\sqrt{2}$ and $\phi=\sqrt{(\phi^0)^2+2\phi^+\phi^-}$.

We consider couplings of $\Sigma_{\text{SM}},\Sigma_{\text{twin}}$ to electrons and twin electrons that explicitly break the symmetry as in \Eq{eq:ExplicitBreaking}. Expanding to quadratic order in the Goldstones gives
\begin{equation}
\begin{aligned}
    \Lag&\supset-\frac{\dme m_e}{2M_{\text{Pl}}^2}\left(|\Sigma_\text{SM}|^2\bar{e}e+|\Sigma_\text{twin}|^2\bar{e}'e'\right)\\
    &=-\frac{ \dme m_e}{8M_{\text{Pl}}^2}\left(\bar{e}e-\bar{e}'e'\right)\phi^+\phi^-+\dots
    \end{aligned}
\end{equation}
Due to the mirror-$\mathbb{Z}_2$ symmetry, the mass correction to $\phi^\pm$ at linear order in the coupling also exactly cancels in this case.
The leading order again arises at quadratic order in the coupling and gives, as for the abelian model, a mass contribution of
\begin{equation}
     \delta m_{\phi^\pm}^2=\frac{(\dme)^2 m_e^2f^2}{4\pi^2 \Mpl^4} \Lambda^2.
\end{equation}
Hence, we find that the quadratic twin ULDM mass is suppressed as in the abelian case, see \Eq{eq:UniversalSuppression}.

Unlike the $U(1)$ model, one of the pNGBs is massless in the $U(2)$ model. 
Indeed, $\phi^0$ does not pick up a mass from the explicit breaking due to a residual $U(1)$ symmetry associated with the $T^3$ generator.

\section{Phenomenology}\label{sec:Pheno}

In this section, we discuss some phenomenological aspects of quadratic twin ULDM, with a focus on its detection prospects on Earth.

First, we note that SM environmental effects break the mirror-$\mathbb{Z}_2$ and can therefore spoil the mass-protection mechanism. 
Given that the quadratic twin ULDM couples non-derivatively to SM operators, these operators modify the pNGB mass in backgrounds where ${\braket{\mathcal{O}_\text{SM}}\neq0}$, while ${\braket{\mathcal{O}_\text{twin}}=0}$.

Consider the coupling to electrons in \Eq{eq:ULDM_electron_interaction}. 
Within a finite density of (non-relativistic) electrons $\braket{\bar ee}\simeq\braket{e^{\dagger}e}\equiv n_e$, such as within Earth, the pNGB mass gets shifted by
\begin{equation}
    \left(m_\phi^2\right)^\text{eff}=m_\phi^2+\frac{\dme m_e}{\Mpl^2}n_e,
\end{equation}
where $\dme$ can be positive or negative.
This mass shift within the Earth (or any other compact object with $n_e\neq0$) leads to screening of the DM field amplitude while gradients thereof are generically enhanced. 
This has been studied extensively for light QCD axions \cite{Hees:2018fpg,Banerjee:2025dlo,delCastillo:2025rbr}.
We show the details of the calculations for both signs of the coupling in \App{app:ScreenSource}, taking into account finite momentum boundary conditions.
The bounds sensitive to amplitudes, such as from clocks and interferometers, can be affected by screening in parts of the parameter space.
In \Fig{fig:parspace}, we show the expected bound with solid lines while keeping the Goldstone naturalness bound as dashed lines of the same color.
Similarly, in parts of the parameter space, experiments sensitive to DM field gradients, such as EP tests, are affected.
In \Fig{fig:parspace} we show the bounds from MICROSCOPE for positive and negative $\dme$.
The resonance features are characteristic of negatively coupled quadratic scalars, assuming a perfectly spherical Earth with homogeneous density \cite{Banerjee:2025dlo,delCastillo:2025rbr}, see \App{app:ScreenSource} for further details. 

The natural region of quadratic twin ULDM can be extensively searched for in the future with $\iso{Th}{229}$ nuclear clocks \cite{Kim:2022ype,Caputo:2024doz}.
Assuming a relative sensitivity of $\delta\alpha/\alpha\sim10^{-24}$ and $d_e\sim(\alpha/4\pi)\times\dme$, we show the nuclear clock projection in \Fig{fig:parspace} as gray dashed.
Stochasticity, not taken into account here, may enhance the reach even further, see \eg \cite{Kim:2023pvt}.

In the case of negative coupling, the mass can become tachyonic inside finite-density objects, which can source the scalar field whether or not it constitutes any fraction of the DM.
This has been extensively studied in the context of (light) QCD axions, where it can lead to extremely stringent constraints \cite{Hook:2017psm,Balkin:2020dsr,Zhang:2021mks,Balkin:2022qer,Balkin:2023xtr,Gomez-Banon:2024oux,Kumamoto:2024wjd}.
For instance, \cite{Balkin:2022qer} showed that the presence of a quadratically coupled light QCD axion can change the mass-radius relationship of white dwarfs which is inconsistent with observations.
These constraints arise when the change of the fermion mass due to the sourced field is relatively large, as is the case for QCD axions, and disappear for a coupling to electrons when $\delta m_e/m_e<10^{-3}$ \cite{Balkin:2022qer,Bartnick:2025lbv,Bartnick:2025lbg}. 

Interestingly, for the quadratic twin ULDM, these bounds can be avoided since the mass change is suppressed as $\delta m_e/m_e\sim\dme f^2/\Mpl^2\ll1$ (see \App{app:ScreenSource} for details).
This opens up the possibility of interesting phenomenology where a sourced field exists around Earth.
We leave a detailed analysis of these effects to future work. 

Similarly, quadratic twin ULDM also avoids bounds from Big Bang nucleosynthesis \cite{Bouley:2022eer}, since they rely on relatively large changes to the electron mass.

\section{Conclusions and Discussion}\label{sec:Conclusions}

We have presented a model of naturally ultralight, quadratically coupled scalar dark matter.
Inspired by the twin-Higgs mechanism \cite{Chacko:2005pe}, we considered a pNGB of some spontaneously broken global symmetry $G$. 
We introduced couplings to our SM and a twin sector, related by a mirror-$\mathbb{Z}_2$ symmetry, which explicitly breaks $G$ to $H_\text{SM}\times H_\text{twin}$.

The explicit breaking generates a mass for the pNGBs, which in the absence of the twin mechanism would have rendered the entire experimentally accessible parameter space unnatural. The novelty of our mechanism is that the mirror-$\mathbb Z_2$ symmetry cancels the corrections at linear order in the coupling, and the leading contribution arises at quadratic order.
The quadratic sensitivity to the coupling results in a huge mass suppression compared to the Goldstone naturalness limit. 
We have explicitly demonstrated this effect for $G=U(1)$ and $G=U(2)$, where roughly 25 orders of magnitude open up in the parameter space of fermion couplings for masses of $\mathcal{O}(10^{-20})\,\text{eV}$.

Finally, we want to comment on the cosmological history of quadratic twin ULDM.
While it can, in principle, be produced by misalignment, the quadratic twin ULDM comes with the usual difficulties of the twin-Higgs cosmology, see \eg \cite{Craig:2016lyx,Chacko:2018vss}.
The mirror-$\mathbb{Z}_2$ has to be broken in the early universe since constraints on the effective number of degrees of freedom do not allow reheating the twin SM.
Therefore, the thermal mass induced by the SM bath is not protected by the mirror-$\mathbb{Z}_2$, which can lead to either larger, smaller, or even tachyonic masses during the early Universe evolution, depending on the sign of the coupling. The resulting cosmological history can have rich phenomenology, with different regimes for ULDM equation of state and potential phase transitions. These consideration might be important for possible production mechanism of quadratic twin ULDM.

\begin{acknowledgements}

We would like to thank Abhishek Banerjee, Kai Bartnick, Zackaria Chacko, Anson Hook, Oleksii Matsedonskyi, Stefan Stelzl and Jure Zupan for fruitful discussions.
We would especially like to thank Anson Hook for pointing out the non-standard cosmology due to thermal mass corrections and Zackaria Chacko for comments on the draft. 
The work of CD is supported by the CNRS IRP NewSpec. MG and ZHA are supported by the Israel Science Foundation under Grant No. 
1424/23. MG is also supported by the US-Israeli BSF grant 2018236 and the NSF-BSF grant 2021779. GP is supported by the ISF, NSF-BSF, and the Minerva foundation. 
KS is supported by a research grant from Mr. and Mrs. George Zbeda and by the Minerva foundation.

\end{acknowledgements}

\appendix

\section{Goldstone naturalness}\label{app:naturalness}

The Goldstone naturalness bounds on the scalar mass from coupling to fermions is
\begin{equation}\label{eq:naiveNaturalness}
    m_\phi^2\gtrsim\delta m_\phi^2\simeq\frac{d^{(2)}_{m_\psi}m_\psi^2\Lambda^2}{32\pi^2\Mpl^2}.
\end{equation}

If we assume that the scalar is a pNGB, we also require naturalness of the quartic $\lambda\sim m_\phi^2/f^2$, where $f$ is the compact field range of $\phi$.
The fermion interaction introduces corrections
\begin{equation}\label{eq:QuartNat}
    \delta\lambda\simeq \frac{(d^{(2)}_{m_\psi})^2m_\psi^2\Lambda^2}{64\pi^2\Mpl^4}\lesssim \frac{\delta m_\phi^2}{f^2}\implies f\lesssim \sqrt{\frac{2}{d^{(2)}_{m_\psi}}}\Mpl.
\end{equation}
This naturalness criterion can be more restrictive than \Eq{eq:naiveNaturalness}, and explains the kink at $\dme\simeq10^{2}$ and $m_\phi\simeq10^{-9}$eV in \Fig{fig:parspace}.

\section{Screening and sourcing} \label{app:ScreenSource}

In this appendix, we discuss the phenomenological implications of a density-dependent scalar mass.
We first consider screening, which suppresses the DM amplitude but enhances gradients, as shown in \cite{Hees:2018fpg,Banerjee:2025dlo,delCastillo:2025rbr}.
Second, we discuss the scenario in which the scalar mass turns tachyonic which leads to sourcing \cite{Hook:2017psm,Balkin:2020dsr,Balkin:2021wea,Balkin:2021zfd,Balkin:2022qer,Balkin:2023xtr,Gomez-Banon:2024oux}.\\

\textbf{Screening}\\

We closely follow \cite{Banerjee:2025dlo} and define momentum \emph{inside} the Earth as $q$ and the momentum \emph{outside} Earth as $k$:
\begin{subequations}
    \begin{align}
        k^2 \equiv \omega^2 - m_\phi^2&\quad\text{(outside)}\\
        q^2 \equiv \omega^2 - m_\phi^2 - \delta m_\phi^2&\quad\text{(inside)}
    \end{align}
\end{subequations}
where
\begin{equation}
    \delta m^2_\phi=\frac{\dme m_e}{\Mpl^2}n_e.
\end{equation}

\begin{figure*}[t!]
    \centering
    \includegraphics[width=0.49\textwidth]{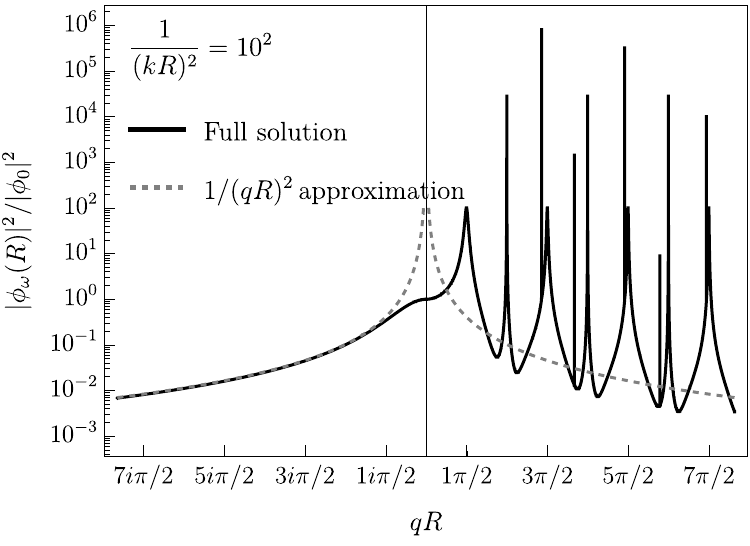}
    \includegraphics[width=0.49\textwidth]{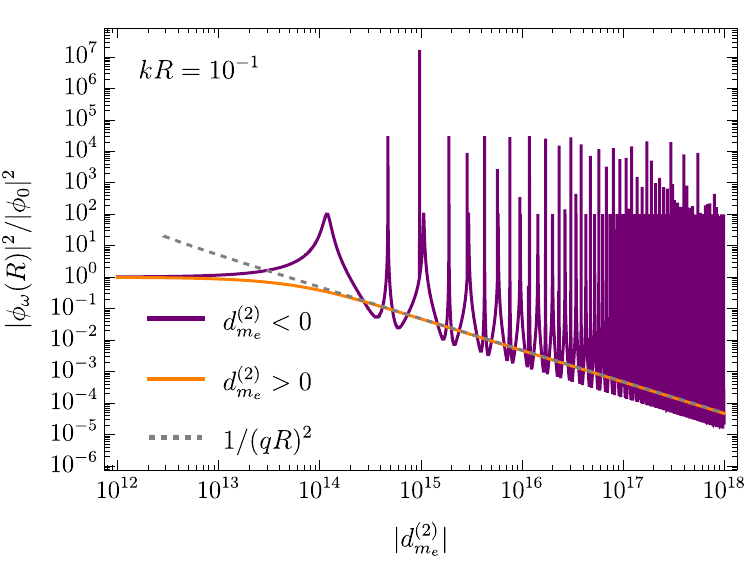}
    \includegraphics[width=0.49\textwidth]{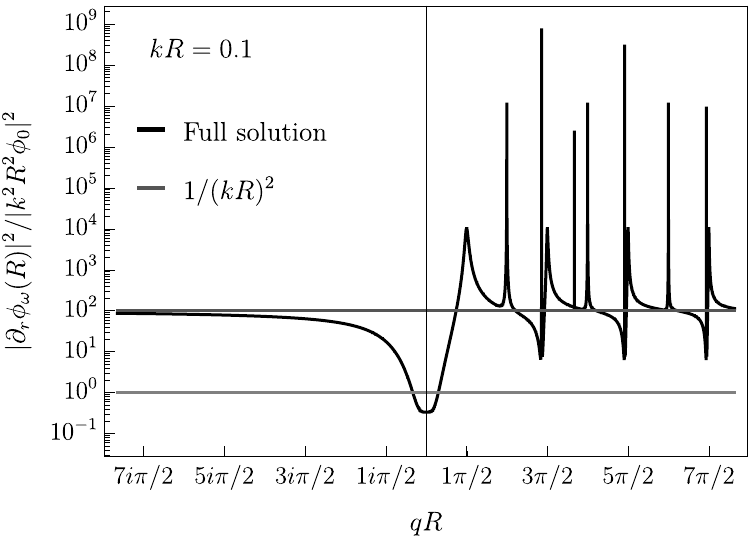}
    \includegraphics[width=0.49\textwidth]{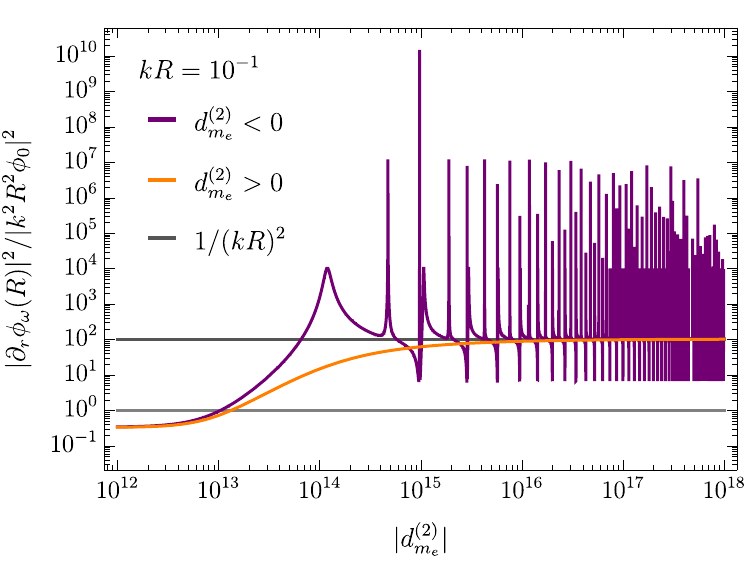}
    \caption{(Left) Dark matter profile and gradient at Earth's surface for increasing ($(qR)^2<0$) and decreasing ($(qR)^2>0$) scalar mass as function of $qR$. 
    (Right) Same but as a function of $\dme$ for both signs.}
    \label{fig:AmpQRplt}
\end{figure*}

The angular averaged steady-state solution to the EOM at the surface of (spherical symmetric, constant density) Earth, is given by
\begin{widetext}
    \begin{equation}
    \begin{aligned}
        \langle |\phi(R)|^2 \rangle_\theta =& \left( \frac{\phi_0}{k^2 R^2} \right)^2 \sum_{\ell=0}^\infty (2\ell + 1) \left| \frac{j_\ell(q R)}{h_{\ell+1}^{(1)}(k R)\, j_\ell(q R) - \left( \frac{q}{k} \right) h_\ell^{(1)}(k R)\, j_{\ell+1}(q R)} \right|^2\\
    \frac{\langle |\hat{r}\cdot\vec{\nabla}\phi(R)|^2 \rangle_{_\theta}}{k^2 \phi_0^2}=&\left(\frac{1}{kR}\right)^6\sum_{\ell=0}^\infty(2\ell+1)\left|\frac{\ell j_\ell(qR)-qRj_{\ell+1}(qR)}{h_{\ell+1}^{(1)}(k R)\, j_\ell(q R) - \left( \frac{q}{k} \right) h_\ell^{(1)}(k R)\, j_{\ell+1}(q R)}\right|^2,
    \end{aligned}
\end{equation}
\end{widetext}
where $\phi_0=\sqrt{\rho_{\text{DM}}}/m_\phi$, $R$ is the radius of Earth, $j_\ell$ and $h_\ell$ are spherical Bessel and Hankel functions.
Depending on the sign of $\dme$, $qR$ increases ($\dme<0$) or decreases ($\dme>0$).
We show the behavior of the dark matter profile at Earth's surface in the upper panels of \Fig{fig:AmpQRplt}.

Most importantly, a decrease of the amplitude can be seen, which is relatively well approximated by $\sim1/|qR|^2$ for $|qR|>\pi/2$. 
In the decreasing mass case ($qR>kR=0.1$ or $\dme<0$) we see deviations from this due to the presence of resonances.
These are, however, not expected to dominate the dark matter profile because we have not included many effects such as an $r$-dependent density profile or drawn different $k$-modes, \eg from a Maxwell-Boltzmann distribution.
Fixing $qR$ corresponds to a choice of the coupling $\dme$, see upper right panel of \Fig{fig:AmpQRplt}.

Let us recap where in the parameter space we expect these effects to be important, see \cite{Banerjee:2025dlo}.
As can be seen in the upper left panel of \Fig{fig:AmpQRplt}, the mass change needs to be significant (for both signs of the coupling) in order for an appreciable effect to occur.
Two conditions must be satisfied for the effect to occur which are $qR\gtrsim\mathcal{O}(1)$ and $|q^2|>|k^2|$, which can both be approximated as $\delta m_\phi R>1$ and $|(\delta m_\phi R)^2|>(kR)^2$, with $\delta m_\phi$ the absolute magnitude of the density induced mass.
These can be roughly interpreted as requiring the object to be large and dense enough, respectively.

\begin{figure*}
    \centering
    \includegraphics[width=0.47\linewidth]{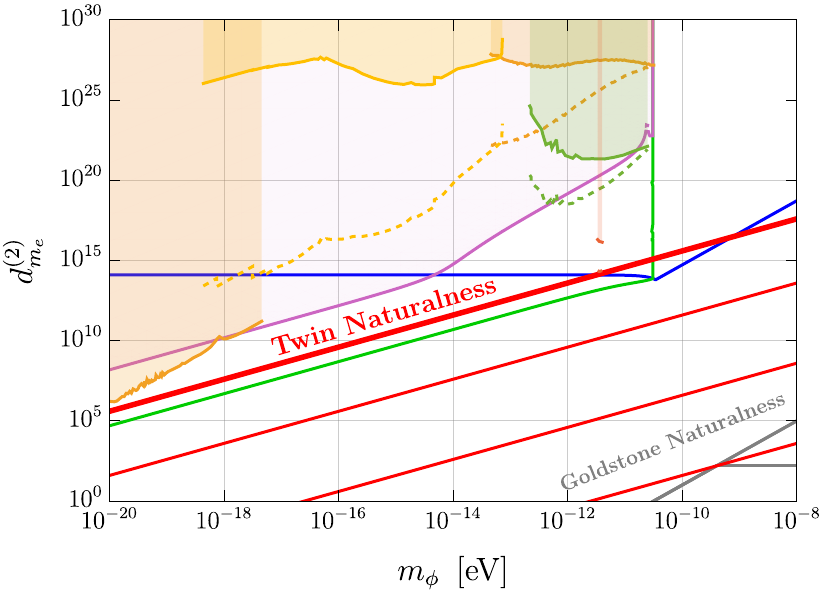}
    \includegraphics[width=0.47\linewidth]{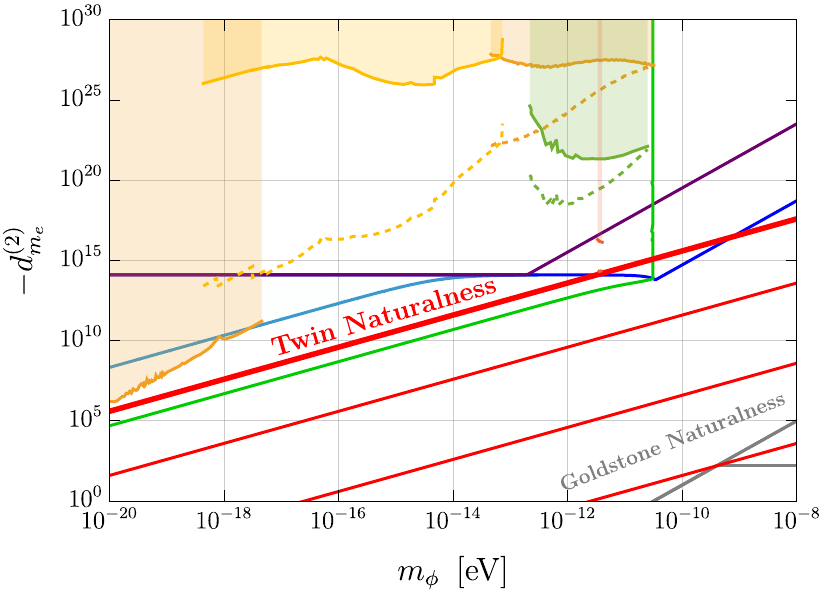}
    \caption{Same as \Fig{fig:parspace} but including regions where for both signs ($\dme>0$ in the left panel and $\dme<0$ in the right panel) the DM field amplitude is screened (above blue line), DM field gradients are enhanced (above green line) and where the quadratic twin ULDM is sourced within Earth (region above the dark purple line in the right panel for $\dme<0$). In the right panel the light blue line is the bound from MICROSCOPE (\cite{Banerjee:2025dlo,Gue:2025nxq}), which is not valid within the dark purple region in which the field is sourced. In this region the effective mass is larger than the vacuum mass and positive since one expands around the finite density minimum.}
    \label{fig:Bounds1}
\end{figure*}

\begin{figure}[t!]
    \centering
    \includegraphics[width=0.4\textwidth]{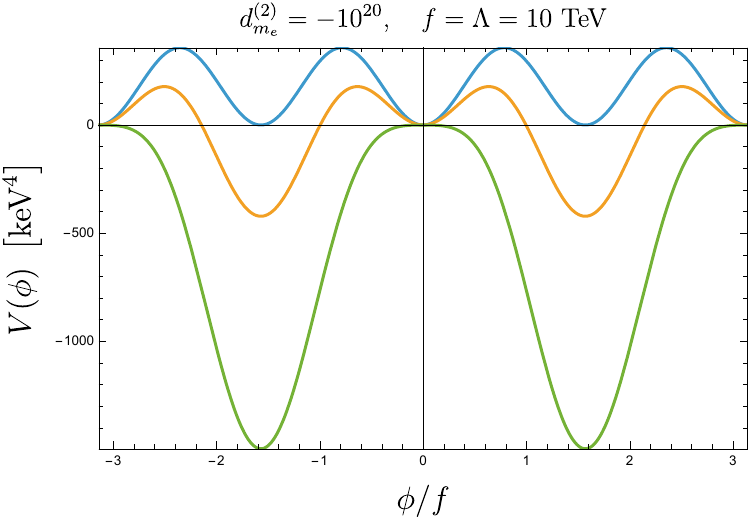}
    \caption{Scalar potential at zero electron density $n_e=0$ (blue), sub-critical density $n_e<n_e^c$ (orange) and super-critical density $n_e>n_e^c$ (green).
    As explained in the text, the finite density contribution to the potential destabilizes the minimum at $\phi_0=0$ for super-critical densities and new minima show up at $\phi=\pm\pi f/2$.}
    \label{fig:PotNegDme}
\end{figure}

Additionally, there is a region where gradients are enhanced.
This can be estimated in the limit where $kR<1$, where the gradient scales as $\partial_r \phi /(k \phi_0)\sim(qR)^2/(3kR)\sim(\delta m_\phi R)^2/(3kR)>1$.
We show these regions for positive (left panel) and negative (right panel) of $\dme$ in \Fig{fig:Bounds1}: 
Above the blue line marks the region in which the amplitude is suppressed, while the region within the green line marks the region where the gradient is enhanced. 
The dark purple line corresponds to the region where the scalar mass is tachyonic, and the solutions presented in \cite{Banerjee:2025dlo} are no longer valid. 
This is because the field is displaced from its in-vacuum minimum and stabilized at $\phi_n=\pi f/2$, see \Fig{fig:PotNegDme}.
We leave the evaluation of the DM profile in this case for future work, see also discussion below.

We want to investigate how current bounds are affected by these effects. 
We first focus on searches sensitive to the amplitude, such as clock and interferometer searches, and recast them using the $\sim 1/(qR)^2$ suppression for both the negative and the positive sign of $\dme$ since we expect the effect of resonances to wash out when realistically accounted for.
We show the recast bounds in \Fig{fig:parspace} and \Fig{fig:Bounds1}.
We recast the bounds obtained for a linear coupling of the form
\begin{equation}
    \frac{d_{m_e}^{(1)} m_e}{\Mpl}\phi \bar{e}e
\end{equation}
by comparing the change in electron mass
\begin{widetext}
\begin{equation}
\begin{aligned}
    \left(\frac{\delta m_e}{m_e}\right)_{\text{lin}}&=\frac{d_{m_e}^{(1)}}{\Mpl}\phi_0\cos(m_\phi t)=\frac{d_{m_e}^{(1)}}{\Mpl}\frac{\sqrt{2\rho_{\text{DM}}}}{m_\phi}\cos(m_\phi t),\\
    \left(\frac{\delta m_e}{m_e}\right)_{\text{quad}}&=\frac{\dme}{2\Mpl^2}\phi_0^2\frac{1}{2}\cos\left(2m_\phi t\right)=\frac{\dme}{4\Mpl^2}\left(\frac{\sqrt{2\rho_{\text{DM}}}}{m_\phi}\right)^2\cos\left(2m_\phi t\right).
\end{aligned}
\end{equation}
\end{widetext}
Therefore, to find the bound on the quadratic coupling we can make the replacements
\begin{equation}
\begin{aligned}
    m_{\phi,\rm quad}&=\frac{m_{\phi,\rm lin}}{2},\\
     d_{m_e}^{(2)}&=d_{m_e}^{(1)} 
     \sqrt{\frac{2}{\rho_{\rm DM}}}\Mpl m_{\phi,\rm quad}.
\end{aligned}
\end{equation}
Taking into account the suppression due to screening, see upper panels of \Fig{fig:AmpQRplt}, implies shifting the amplitude by a suppression factor of 
\begin{equation}
    \phi_0^2\to\frac{1}{\lvert(qR)^2\rvert}\phi_0^2=\frac{1}{\lvert(v_0 m_\phi R)^2\pm\frac{|\dme|}{4\Mpl^2}m_e n_e R^2\rvert}\phi_0^2,
\end{equation}
which enters the effective quadratic coupling.
The bound obtained without screening is shown in \Fig{fig:parspace} and \Fig{fig:Bounds1} in dashed lines, while the screening-accounted bound is shown in solid lines.
Notice that we focus on the regime where $kR<1$, which corresponds to $m_\phi\lesssim10^{-11}\,\text{eV}$, since reliable calculations can only be done in this regime \cite{Banerjee:2025dlo}. 
Therefore, we are not showing bounds from~\cite{Antypas:2019qji} and~\cite{Savalle:2020vgz} which lie in the $m_\phi>10^{-11}\,\text{eV}$ regime. 

Contrary to the reduced sensitivity of amplitude-dependent experiments, gradient-dependent experiments would generically see an enhancement $\sim 1/(kR)^2$.
This fact has recently been exploited for EP tests such as the MICROSCOPE experiment which is sensitive to $\sim\phi\nabla \phi$ in \cite{Gue:2025nxq} for the axion coupling, using the profiles calculated in \cite{Banerjee:2025dlo}.

The analysis is straightforward.
The E\"otv\"os parameter for MICROSCOPE for quadratically coupled dark matter is given by
\begin{equation}
    \eta=2\frac{|\vec{a}_A-\vec{a}_B|}{|\vec{a}_A+\vec{a}_B|}=(Q_A^M-Q_B^M)\frac{\Mpl^2R_\text{mic}^2}{M_\oplus}\frac{\dme }{\Mpl^2}\left\lvert\phi_{\text{out}}\vec{\nabla} \phi_{\text{out}}\right\rvert,
\end{equation}
where $\phi_{\text{out}}$ is a function of $m_\phi R_\oplus,m_\phi R_\text{mic}$ and $m_\phi(R_\text{mic}- R_\oplus)$ \cite{Banerjee:2025dlo} and $Q_i^M$ are the dilatonic charges.
Here we use $R_\text{mic}\simeq710\,\rm km$.
For the electron coupling, the dilatonic charges for $\dme$ are given by \cite{Damour:2010rp}

\begin{equation}
    Q_i^M=F_A\left[5.5\times10^{-4}\frac{Z}{A}\right].
\end{equation}

\begin{figure*}[t!]
    \centering
    \includegraphics[width=0.49\linewidth]{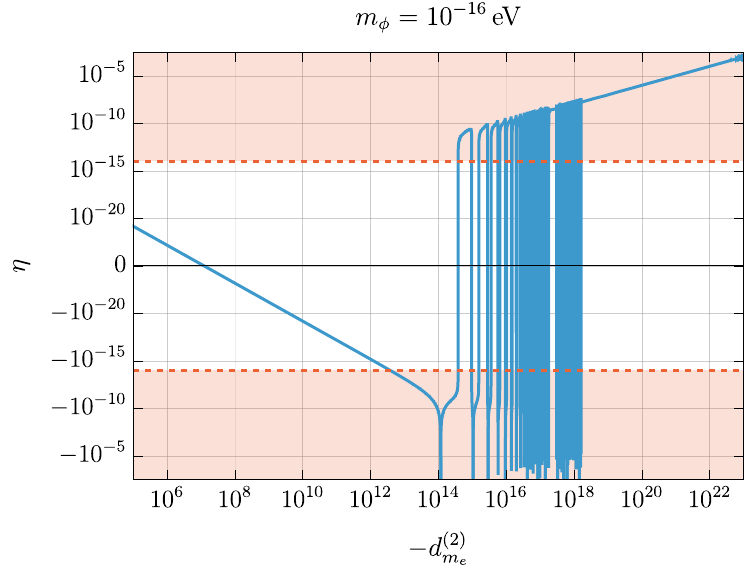}
    \includegraphics[width=0.49\linewidth]{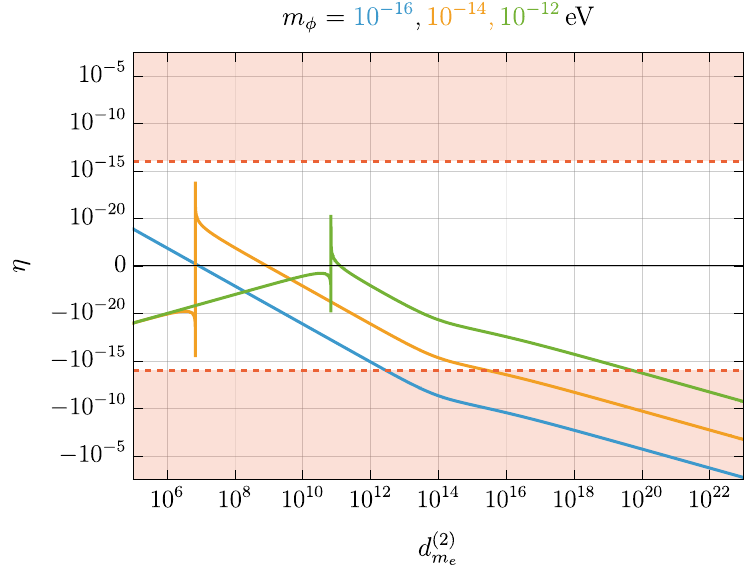}
    \caption{E\"otv\"os parameter as a function of negative (left) and positive (right) $\dme$. 
    The red region is excluded by MICROSCOPE.
    }
    \label{fig:EotvosPar}
\end{figure*}

In the case that $kR_\oplus<1$ it is sufficient to only consider the $\ell =0$ mode.
The bound from Microscope is roughly $|\eta|<10^{-14}$ and 
plot the E\"otv\"os parameter as a function of $\dme$ in \Fig{fig:EotvosPar}.

This translates to the bounds on $\dme>0$ (purple) in the left panel, and $\dme<0$ (yellow) in the right panel of \Fig{fig:parspace} and \Fig{fig:Bounds1}.
We would also like to mention that the resonances that occur for $\dme<0$ are likely due to the assumption of a spherically symmetric homogenous sphere and that they should disappear once some realistic profile is taken into account.\\

\textbf{Sourcing}\\

Sourcing can occur if $\dme<0$. Within a finite density background, the mass is reduced and at large enough densities it becomes tachyonic.  
This happens at densities
\begin{equation}\label{eq:DenseEnough}
    n_e>n_e^c\equiv-\frac{m_\phi^2 \Mpl^2}{\dme m_e}\gtrsim -\frac{\dme m_e f^2 \Lambda^2}{4\pi^2 \Mpl^2}.
\end{equation}
At these densities the point $\phi_0=0$ ceases to be a minimum and new minima appear at $\phi_0^n=\pm\pi f/2$.
In \Fig{fig:PotNegDme} we show the non-relativistic 1-loop Coleman-Weinberg potential for $n_e=0$ (blue) as well as sub- (orange) and super-critical densities (green).
As we can see, the minimum at $\phi=0$ gets destabilized and the minima at $\pm \pi f/2$ are energetically preferred.
Therefore, the electron mass, which depends on the $\phi$ VEV is now reduced by
\begin{equation}
\begin{aligned}
        m_e(\phi)=&\,m_e\left[1+\dme\frac{f^2}{2\Mpl^2}\sin^2\left(\frac{\phi}{f}\right)\right],\\
        m_e(\phi_0^n)=&\,m_e\left[1+\dme\frac{f^2}{2\Mpl^2}\right]<m_e.
\end{aligned}
\end{equation}

This, however, can only happen if the object is both dense and large enough. 
Assuming that the mass inside is dense enough \Eq{eq:DenseEnough}, sourcing occurs if also
\begin{equation}\label{eq:LargeEnough}
    R\gtrsim\frac{\Mpl}{\sqrt{\dme}}\frac{1}{\sqrt{n_e m_e}}.
\end{equation}

Note that no linear coupling is generated in regions of the object where the field sits in the finite density minimum of the potential.
If gradient effects are important the scalar can be between the two minima within regions of the star and linear couplings would be induced.
These would give rise to strong bounds from \eg stellar cooling \cite{Bartnick:2025lbv}, which we do not consider here.

Even without linear couplings, sourcing can lead to large observable effects.
In \cite{Balkin:2022qer,Bartnick:2025lbv,Bartnick:2025lbg} it is shown that strong bounds can be derived if the mass reduction of the electron is large enough.
Concretely, that implies that the bound from mass-radius observations of White Dwarfs disappears for mass changes smaller than $\delta m_e/m_e<10^{-3}$.
In our case, if we identify the field range with the cutoff $f=\Lambda= 10\,\text{TeV}$, we find that the electron mass change is tiny except for the largest values of $\dme$, and therefore we do not show bounds from modified white dwarf mass-radius curves.

\bibliography{bibliography}
\bibliographystyle{apsrev4-1}

\end{document}